# How to build a DNA search engine like Google?

Wang Liang, Fang Bo

Tencent, SOSO, 100080, P.R. China

Huazhong University of Science and Technology,430074, P.R. China

*To whom correspondence should be addressed. E-mail:wangliang.f@gmail.com

[Abstract] This paper proposed a new method to build the large scale DNA sequences search system based on web search engine technology. We give a very brief introduction for the methods used in search engine first. Then how to build a DNA search system like Google is illustrated in detail. Since there is no local alignment process, this system is able to provide the *ms* level search services for billions of DNA sequences in a typical server.

## 1 Introduction

The exponentially increasing of biological data poses new challenges for bioinformatics in the post-genome era. Now if you want to search a DNA sequence in all of current DNA databases, you will find it's a very tough mission. But if you want to search a word sequence in billions of documents in Internet, you only need several *ms* using Google. So could we build a DNA search engine like Google? This paper just discusses this "simple" question.

Now most DNA search and comparing methods are similar to BLAST/FASTA algorithm, which compares one sequence with the other sequences on by one [1,2]. Although many heuristic and pre-index methods could greatly reduce the search time, it's still difficult to meet the challenge in this DNA information explosion period. Many researchers agree that high performance search algorithm is demanding in current research of bioinformatics.

We may find some tips from the history of text information retrieval. If we need search a word sequence in several documents, we could scan each document by some string matching algorithms like KMP. But if there are millions of documents, the search time will become intolerable. So many pre-process methods like Tree based methods are proposed. Many of them are also applied in DNA analysis [3]. But for mass data like Internet information, the inverted index based search systems are almost the only choice [4]. We could use a simple example to show the inverted index methods:

Three documents:

D0= "a big apple";

D1= "a apple I love apple";

D2 = "big pig eat apple";

The "inverted index" is an index data structure storing a mapping from content, such as words or numbers, to its locations in a database file, or in a document or a set of documents. In search engine, the inverted index is normally the mapping from "word" to "document":

| word | Document ID |
|------|-------------|
| a    | D0,D1       |
| big  | D0,D2,      |

| apple | D0,D1,D1,D2 |
|-------|-------------|
| I     | D1          |
| pig   | D2          |
| love  | D1          |

If we want to search "a apple", we first search the words "a" in word column and obtain its corresponding Document ID list: R("a")={D0,D1,D2}. Then for "apple", R("apple")={D0,D1,D1,D2}. The search results is the intersection set of R("a") and R("apple"):

$$R = R("a") \bigcap R("apple") = \{D0, D1\} \bigcap (D0, D1, D1, D2) = \{D0, D1\}$$

The search result is D0 and D1. Normally, a scoring method is designed to rank the results. The simplest method is TF (term frequency) score, which is defined as the same words amount between query and result sequence. Here the score of D0 is 2, D1 is 3. So D1 will rank first and D0 in second.

The common string matching method need scan each target document. Its time complexity is more than O(n), in which n is the number of documents. In some heuristic method, we only need scan some "seeds" of all documents. But its complexity is still related to the number of target documents. But for inverted index system, its time complexity mainly depends on the number of words in query (m). The proper set merging algorithm could ensure the search time complexity is O(NP(m)). Normally the document is billion level, which is much more than O(NP(m)). It's just the "secret" of Google to provide the ms level search service for mass WWW documents.

## 2 DNA search engine design
### 2.1 Brief idea

The inverted index could be used to index any symbol sequence and provide the quick search service, no matter it's music, noise etc. We only need solve two problems. First, how to divide or segment the sequence into "words" ? Second, we should ensure the word set or vocabulary is not too long. In addition，For DNA sequence like full genome, we should also divide the long sequence into short "document" to directly apply the mature technology in search engine. The following parts just discuss these questions.

### 2.2 DNA document

In this paper, we use the full genome as experiment data. We could index the position of words, but we need revise many available search methods. So we divide long sequence into short sequences, just like dividing a long book into short paragraphs. There is no a strict standard to determine the proper length of "document". Normally, this length should be a multiple of the length of most queries. Here assume a common DNA query is about 100 bps. So we use 500 as the length of DNA "document". This length could be adjusted according to the search requirement.

The advantages of "document" could be easily explained. For query "I love apple", the "I really love apple" could be regarded as a matching sequence. But for "I + (1000 words) + love+ (10000 words) + apple" is obviously not a matching sequence. The "document ID" list merging in search process will be much easier than merging operation based on position.

Here the genome of Arabidopsis is successively divided into 500 bps length documents. There are about 272,336 documents. Each is marked by their position in genome, which is as same as the description of FASTA format. We also give a unique document ID to each document to identify them.

**2.3 DNA words and segment**

For some language like English, the sequence could be easily segmented into words according to space and punctuation. But for DNA sequence, there is no space and punctuation. This problem can be solved considering some languages with no natural delimiters like Chinese. The simplest method to index Chinese is using n-grams methods to (normally, 2-grams) segment the sentence. To obtain the n-grams segment, one shifts progressively by one base a "reading window" of length n along the sequence. For example, a sentence:
T="ABCDE";
The 1-gram segment divided the T into {A, B, C, D, E};
The 2-grams segment divided the T into {AB, BC, CD, DE}

For a query "ABC", it's divided into "A"+"B"+"C" (1-gram) or "AB"+"BC" (2-grams) and search them in related inverted files.

Obviously, we could also use the n-grams segments to divide the DNA sequences. The only question is to determine a proper "word" length to ensure the "vocabulary" is not too big. If we select 1 as the length of DNA word, there are only 4 words {A, T, C, G}. The DNA sequences corresponding to one word will too many to process the union operation. According to the practice, the word in vocabulary should not more than 10 million level. So the proper word length of DNA could be 7 to 12.

Here we could select length 12 ($4^{12} = 16,777,216$).

After 12-ngrams segmenting the DNA sequence, we could easily apply the current search engine technology to create inverted index and build a DNA search engine. We write a simple DNA search system based on a tutorial search system [5].

**2.4 DNA search process**

We use the genome of Arabidopsis as experiment data and build a search system. But we find there is no search result for most queries. It's mainly because there is almost no exact match "Document" for query sequence. To deal with this problem, we only need adjust the search process.

First, we sort the query word by the Document Frequency of word. Document Frequency of a word is defined as number of documents containing this word. Then we merge the document list according to this sorted query list successively, until the number of candidate results is less than a threshold.

For example, an inverted index:

| word | Document ID |
|------|-------------|
| W1   | D1          |
| W2   | D0,D2,D3,   |
| W3   | D0,D2,D4    |
| W4   | D0,D2       |

The query is {W1,W3,W4}. If we search W1 first, R("W1")={D1}, Then for R("W3")={ D0,D2,D4}, their union:

$$R = R("W1") \bigcap R("W3") = \{D1\} \bigcap (D0, D2, D4) = \{\}$$

So we could only use {D1} as the candidate result.

But if we rank the query according to their Document frequency: {W3,W4,W1}. Then for R("W3")={ D0,D2,D4}, R("W4")={ D0,D2,D4}, their union:

$$R = R("W3") \bigcap R("W4") = \{D0, D2, D4\} \bigcap (D0, D2) = \{D0, D2\}$$

Although there also no result for:

$$R = R("W3") \bigcap R("W4") \bigcap R("W1") = \{\}$$

But we could use the result in the previous step "{D0,D2}" as the search results. Obviously, this result is better than "{D1}", which only match one word.

**2.5 Compare with BLAST**

Normally, there are two steps in current fast alignment algorithms like BLAST. First,"the search stage", the program quickly detects sequences which are likely to be homologous with query sequence. For example, BLAST segments the query into 11 length "words" and then finds the target sequences matching at least one "word" as candidate sequence. Second, "the local alignment stage", program examines the candidate sequence in more detail and produces alignments for the sequences which are indeed homologous according to some criteria. For BLAST, it applies S-W method to rank the candidate sequences and get the final "search results".

But for our system, there is only one step, "the search stage". Our programs also 12-segment the query and find the sequence matching one word as candidate sequence. But in the next step, we continue to find sequence matching more words in candidate sequences of previous step. This process is repeated until the number of candidate sequence is less than a threshold or the exact match sequences are found. We finally get a results list sorted by TF scores.

The TF score is a different similarity definition. This scoring method and its improvement like TF*IDF are all designed based on Statistical Language Models. It's difficult to compare the search engine methods and BLAST in theory in a short paper. So we only describe their differences in using.

For genome of Arabidopsis thaliana, many short queries only return the "No significant similarity found" (blast(n) in NCBI). But our system could give the search results for almost all queries. Half a loaf is better than no bread. For the matching sequence of BLAST，most of them also appear in the in top the result list of our system, especially for the results with few gaps. But few results obtain very low ranking scores in our system.

The main advantage of DNA search engine is that it could provide much faster search speed than BLAST. Only for our theory proof system, it could provide 10 query/second search service and return the results in <200ms for full genome of Arabidopsis thaliana. This service run in an AWS micro instance(1 cpu, 600M memory). Based on the mature search system like Lucene, we could easily index billions of DNA sequence and provide >100 search services per-second by a single common server. So this system may be a good choice for the preliminary search of a DNA sequence.

**2.6 Further improvements**

Although the 2-grams segment is also applied in some simple Chinese search system, most large Chinese search engine all use the "vocabulary based segment". We could use a simple example to show the advantages of "vocabulary based segment".

For example:

A document D1="ABIGAPPLE"; its 3-grams segments: {ABI, BIG, GAP, GPI,APP,PPL,PLE}

For query Q="AAPPLE", its 3-grams segments:{AAP,APP,PPL,PLE}. The D1 will not return as the search results of Q.

But if we use the "vocabulary based segment". For D1, its segments {A, BIG,APPLE },for Q,{A,APPLE}. So the D1 will return as the search result of Q.

Moreover, the words {ABI, IGP, GPI…} are not a real meaningful "word", we needn't index them. We only need index {A, BIG, APPLE}. This could reduce the size of inverted files. It's very important to in building the large scale search system.

To apply the "vocabulary based segment" method, we should build a "DNA vocabulary" containing different length "words", and then segment the DNA sequence properly. In fact, there have been many researches for these two topic in Chinese search engine and n-grams languages model research area. An interesting result is that the "2-grams" has similar search effects to "vocabulary based segment" in Chinese search engine. So we will discuss the "vocabulary of DNA" in another paper.

# 3 Conclusions

This paper gives a solution to build a large scale DNA search engine. It could provide the quickly search service for DNA sequence. This method could also be easily extended to search protein sequence. Now search engine has become a very mature technology. Based on the method in this paper, almost all the other search engine technologies like distributed search methods could be directly applied in DNA analyzing.